\documentclass{aa}
\usepackage{graphics}

\newcommand{\kms}   {~km~s$^{-1}$}

\newcommand{\cmd}   {~cm$^{-2}$}
\newcommand{\cmt}   {~cm$^{-3}$}

\newcommand{\vlsr}  {$v_{\rm LSR}$}

\newcommand{\et}    {et al.}

\newcommand{\iras} {IRAS~08194$-$4925}

\newcommand{\hco}   {HCO$^+$}
\newcommand{\htco}  {H$^{13}$CO$^+$}
\newcommand{\met}   {CH$_3$OH}
\newcommand{\JA}    {\mbox{2$_0$--1$_0$~A$^+$}}
\newcommand{\JE}    {\mbox{2$_{-1}$--1$_{-1}$~E}}
\newcommand{\JEo}   {\mbox{2$_0$--1$_0$~E}}
\newcommand{\J}[2]  {\mbox{#1--#2}}
\newcommand{\JK}[4] {\mbox{#1$_{#2}$--#3$_{#4}$}}

\newcommand{\arcdeg}{\mbox{$^\circ$}} 
\newcommand{\nodata}{ ~$\cdots$~ }    

\begin{document}


\title{The origin of the molecular emission around the southern hemisphere
Re~4 IRS -- HH 188 region}

\author{
Josep Miquel       Girart\inst{1}
\and
Serena Viti\inst{2}
}

\offprints{J. M. Girart: girart@ieec.cat}

\institute{
Institut de Ci\`encies de l'Espai (CSIC- IEEC), 
Campus UAB - Facultat de Ci\`encies, Torre C5 - parell 2, 08193 Bellaterra, 
Catalunya, Spain
\and
Department of Physics and Astronomy, University College London,
London, WC1E 6BT, England
}

\date{Received ...; accepted ...}


\abstract
{
We present SEST observations of the molecular environment ahead of the southern
Herbig-Haro object 188 (HH~188), associated with the low-mass protostar Re~4
IRS.   We have also used the SuperCosmos H$\alpha$ survey to search for
H$\alpha$ emission associated with the Re~4 IRS -- HH~188 region.
}
{The aim of the present work is to study the properties of the molecular gas 
and to better characterize this  southern star forming region.
}
{We mapped the HCO$^+$ 3--2 and H$^{13}$CO$^+$ 1--0 emission around the YSO and
took spectra of the CH$_3$OH \JA\ and \JE\ and SO \JK{6}{5}{5}{4}\ towards  the
central source. Column densities are derived and different scenarios are
considered to explain the origin of the molecular emission. 
} 
{HCO$^+$ arises from a relatively compact region around the YSO; however, its
peak emission is displaced to the south following the outflow direction.  Our
chemical analysis indicates that a plausible scenario is  that most of the
emission arises from the cold, illuminated dense gas ahead of the HH~188
object.  We have also found that HH~188, a high excitation object, seems to be 
part of a parsec scale and highly collimated  HH system. Re~4 IRS is probably a
binary protostellar system, in the late Class 0 or Class I  phase. One of the
protostars, invisible in the near-IR, seems to power the HH~188 system. }{}
\keywords{
ISM: individual: HH~188 --- 
ISM: abundances --- 
ISM: clouds ---
ISM: molecules ---
Radio lines: ISM ---
Stars: formation 
}

\authorrunning{Girart \& Viti}
\titlerunning{A study of the Re~4 IRS/HH 188 region}
\maketitle

\section{Introduction}

Re~4 is an optical reflection nebula, also classified as a cometary globule
(Reipurth \cite{Reipurth81}), in the dark cloud listed as No. 109 by Sandqvist
(\cite{Sandqvist77}). This cloud is located within the Gum Nebula at a distance
of $\sim450$~pc. Spectroscopical observations in the optical of the nebula 
(Graham \cite{Graham86}) shows a very red continuum emission and some
absorption (Na~I~D) and emission lines, mainly $[$SII$]$, H$\alpha$, $[$OI$]$.
The emission line strengths resemble those  of the low-excitation HH objects. 
Re~4 is associated with a low--mass protostar.  Emission associated with the
protostar or its dense circumstellar environment is  detected from near--IR
(2MASS~J08205824-4934456), mid and far IR (\iras ) through to millimeter
wavelengths (Graham \cite{Graham86}; Reipurth et al.  \cite{Reipurth93};  Dent
et al. \cite{Dent98}). The protostar (hereafter Re~4 IRS)  has a bolometric
luminosity of   $\sim$30 L$_{\odot}$  (Cohen \& Schwartz  \cite{Cohen87}). 
There is a star located only $3\farcs1$ northeast of Re 4 and detected also in 
the near--IR (2MASS~J08205862-4934443). Its optical spectroscopic profile 
suggests that  is a K5 V foreground star (Graham \cite{Graham86}).  
Re~4 IRS is associated with a Herbig-Haro object,  HH~188, 
which appears as a series of few knots extending about 10$''$ south of the Re 4
nebulosity (Graham \cite{Graham86}).  The optical spectra of these knots show 
emission in $[$SII$]$, H$\alpha$, $[$OI$]$ and $[$NaII$]$.

VLT near--IR observations show that the Re~4 reflection nebulosity 
extends  further away from the protostar  with a S--like morphology, 
blue in the  south, red
in the north, probably indicating that the northern lobe is further  from us
and more embedded in the parent cloud (Zinnecker et al.  \cite{Zinnecker99}).
This peculiar shape is possibly tracing excavated cavities by the outflow
associated with HH~188. The VLT images show hints of  what appear to be a
binary jet emanating from the obscured centre, with an  offset angle between
the two jets of 5--10$\arcdeg$  (Zinnecker et al. \cite{Zinnecker99}).

In this paper we present SEST observations of the extended molecular emission
around Re~4 IRS and HH~188 as well as complementary data taken from public
optical/near-IR data.  In \S~2 we describe the observations. In \S~3 we
describe the results of the observations and we analyze the molecular emission
around Re~4 and HH~188.    In \S~4 we describe the search for possible new HH
objects associated with HH~188 by using the H$\alpha$ SuperCosmos survey. In
\S~5 we discuss the properties of Re~4 IRS,  HH~188 and its association with a
larger HH system. We also discuss the different possibilities  for the origin
of the molecular emission. In \S~6 we summarize our findings.

\section{Observations}

The observations were carried out in January 10 and 11 2003 with the SEST 15m
radio telescope at La Silla (Chile). We used the dual receiver capability of
the telescope to observe two molecular lines simultaneously, one at 3~mm and
the other at 1~mm: hence we observed simultaneously the \hco\ \J{3}{2}\ and
\htco\ \J{1}{0} as well as the SO \JK{6}{5}{5}{4} and \met\ \JA. We used the 
High Resolution Acousto-Optic Spectrometer, which provided a spectral resolution 
of $\sim$0.14 and 0.05~\kms\ at 3 and 1~mm, respectively (the 1~mm data was
smoothed to a resolution of $\sim 0.3$~\kms\ to increase the signal-to-noise
ratio).  The observations were taken using the frequency switching mode, which
gives a bandwidth of 120 and 30~\kms\ at 3 and 1~mm, respectively . For the
\hco\ and \htco\ lines we did a grid of 26 points, with a cellsize of 20$''$,
covering an area of $1\farcm5\times1\farcm5$ around Re~4 IRS. The SO and
\met\ were only observed towards the IR source.  Table~\ref{tsest} lists the
lines observed, their frequencies, the SEST FWHM (full width at half maximum) 
beam size for these
frequencies, the spectral resolution of the data and the $RMS$ noise in units
of main beam temperature. Average system temperatures during the observations
were about 150, 180, 170, 600~K at 86, 97, 220 and 267~GHz, respectively. Main
beam efficiency at these same frequencies were 0.75, 0.72, 0.52 and 0.42.

%
     \begin{table}
     \caption[]{Lines detected with the SEST telescope}
     \label{tsest}
     \[
\begin{tabular}{l@{\hspace{0.1cm}}c@{\hspace{0.10cm}}r@{\hspace{0.2cm}}c@{\hspace{0.1cm}}c@{\hspace{0.1cm}}r}
     \noalign{\smallskip}
     \hline
     \noalign{\smallskip}   
\multicolumn{4}{c}{} &
\multicolumn{1}{c}{$\!\!\!\!\Delta v$} & 
\multicolumn{1}{c}{$\!\!\!\!\!\!T_{\rm mb}$} 
\\
\multicolumn{2}{c}{} &
\multicolumn{1}{c}{$\nu$} &
\multicolumn{1}{c}{$\!\!\!\!\!\!\!$Beam} & 
\multicolumn{1}{c}{$\!\!\!\!$(km$/$} &
\multicolumn{1}{c}{$\!\!\!\!\!\!rms$} 
\\
\multicolumn{1}{c}{Molecule} &
\multicolumn{1}{c}{Transition} &
\multicolumn{1}{c}{(GHz)} &
\multicolumn{1}{c}{(FWHM)} &
\multicolumn{1}{c}{$\!\!\!\!$s)} &
\multicolumn{1}{c}{$\!\!\!\!\!\!$mK} 
\\
     \noalign{\smallskip}
     \hline
     \noalign{\smallskip}     
\htco & \J{1}{0}        & 86.7543 & 57$''$ & 0.15& 46 \\
\met  & \JA$^a$         & 96.7414 & 51$''$ & 0.13& 38 \\
SO    & \JK{6}{5}{5}{4} &219.9494 & 24$''$ & 0.12& 54 \\
\hco  & \J{3}{2} &267.5576 & 20$''$ & 0.19& 110 \\
     \noalign{\smallskip}
     \hline
     \end{tabular}
     \]
     \begin{list}{}{}
     \item[$^{a}$] These spectra included also the \JK{2}{-1}{1}{-1} E line.
     \end{list}
    \end{table}
%

     \begin{figure} 
     \resizebox{8cm}{!}{\includegraphics{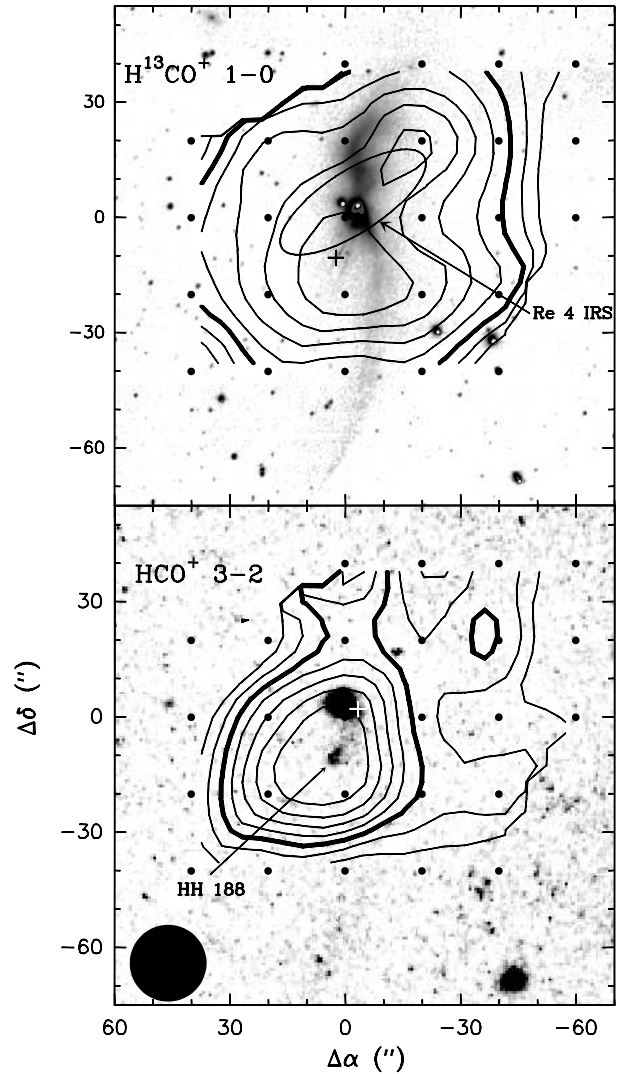}}
     \hfill
     \caption[]{     
     {\em Top panel: }
Superposition of the K$_s$ image (from Zinnecker et al.  \cite{Zinnecker99})
with the SEST contour map of the \htco\ 1--0 emission towards the 
HH~188 region. 
     {\em Bottom panel: }
Superposition of the H$\alpha$ image (from the 
AAO/UKST SuperCOSMOS  H$\alpha$
survey, Parker et al. \cite{Parker05}) with the SEST contour map  of the \hco\
3--2 emission.  
For both  lines the emission was averaged over  the 2.75--4.80~\kms\ \vlsr\ 
interval.  The contour levels are 0.15 to 0.5 with 0.05 K~\kms\ steps for
\htco\  1--0, and 0.3 to 1.0 with 0.1  K~\kms\ steps for \hco\ 3--2.  The
center of the maps corresponds to the position  $\alpha$(J2000)$=08^h 20^m
58\fs59$ and  $\delta$(J2000)$=-49^{\circ} 34' 47\farcs8$. Small filled circles
show the  observed positions. The ellipsoid shows the position uncertainty of
IRAS 08194$-$4925. The black cross in the top panel shows the position  of
HH~188. The white cross in the bottom panel shows the position of  Re~4 IRS
derived from the VLT Ks image. The filled circle at the  bottom--left corner
of the bottom panel shows the beam size of  the \hco\J{3}{2} map, $20''$. 
}
     \label{fmap} 
     \end{figure}
%
%
     \begin{figure} 
     \resizebox{8cm}{!}{\includegraphics{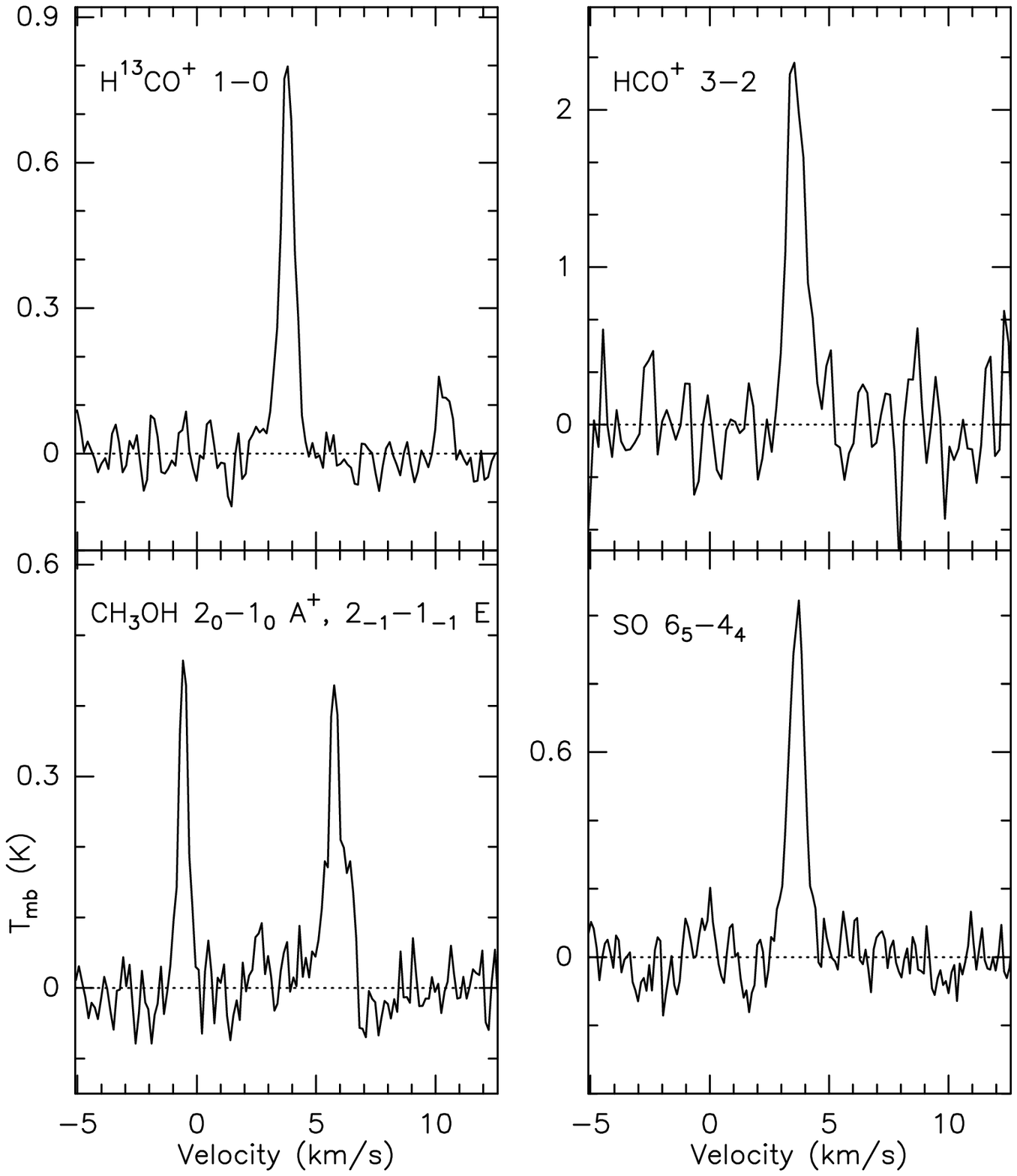}}
     \hfill
     \caption[]{     
     SEST spectra at the ($0''$,$0''$) of the \hco\ \J{3}{2}, \htco\J{1}{0}, 
\met\ \JA\ and \JE, and  SO \JK{6}{5}{5}{4} lines.}
     \label{fcenter} 
     \end{figure}
%

\section{Results and Analysis}
 
Figure~\ref{fmap} shows the integrated emission map of the \hco\ \J{3}{2}\ and
\htco\ \J{1}{0}\ transitions. Figure~\ref{fcenter} shows the spectra of the
\hco\  isotopes as well as those of the SO and \met\ lines.  
Table~\ref{tb:obs} shows the results of Gaussian fits to the spectra at four
different positions. The emission  appears to be relatively compact around
Re~4 IRS.   From the \hco\ \J{3}{2}\ map, which has the highest angular
resolution ($\sim 20''$), the FWHM contour has a diameter of $\sim50''$, which
implies a deconvolved size of $\sim 45''$ or 0.09 pc (assuming that the
emission arises from a single gaussian component).   The peak of the  HCO$^+$ 
\J{3}{2}\ emission is near HH~188, approximately at the $+5'', -12''$  offset
position, i. e. $\alpha$(J2000)$=08^h 20^m 59\fs2$ and
$\delta$(J2000)$=-49^\circ 35' 00''$.

\begin{figure} 
     \resizebox{8cm}{!}{\includegraphics{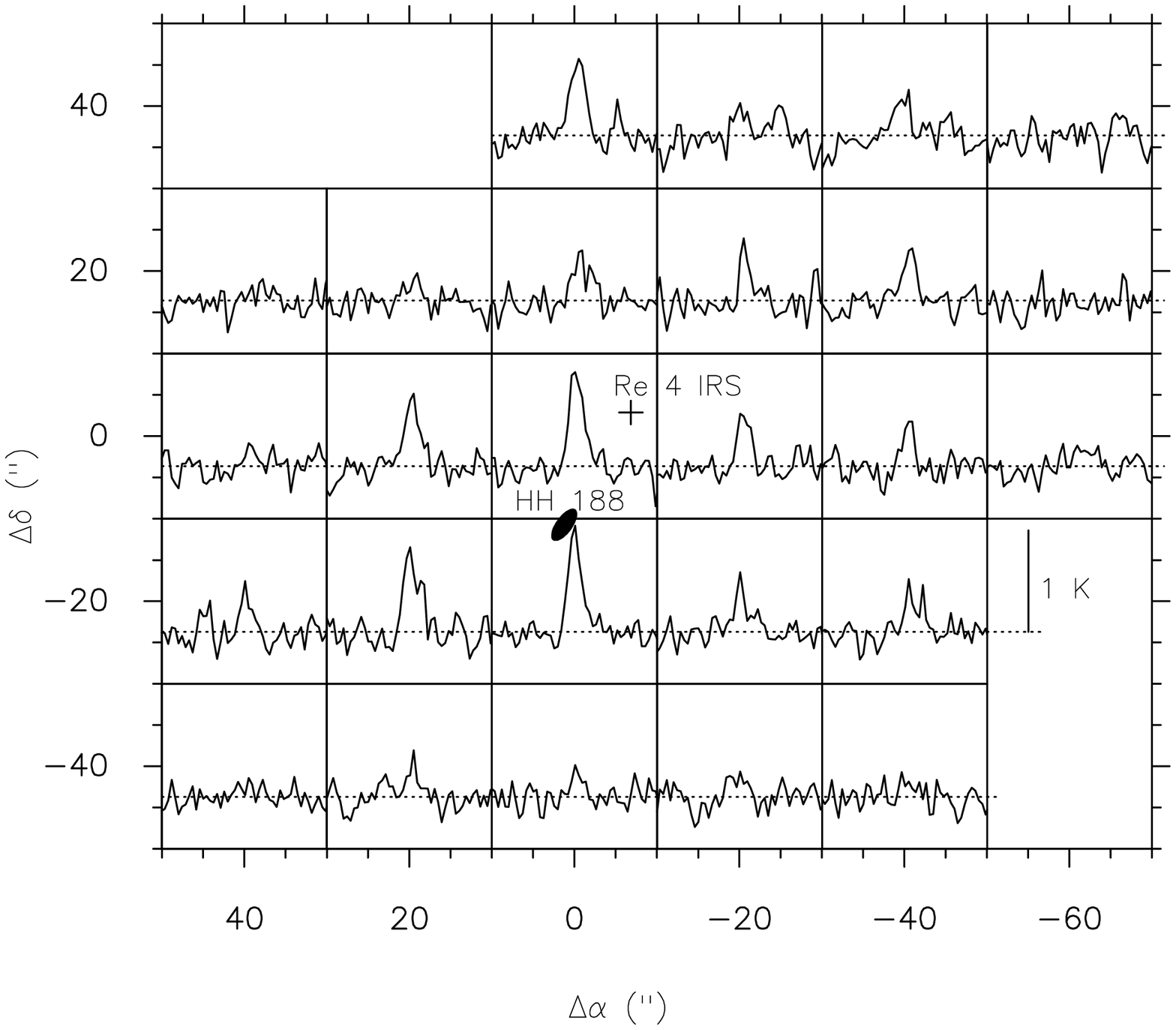}}
     \hfill
     \caption[]{    
Spectra of the \hco\ \J{3}{2}\ towards HH~188. The velocity range for  all the
spectra is from $-1$ to 8~\kms, and the main brightness temperature range
(abscissa axis) is from $-0.5$ to 1.15~K. The scale for 1 K is shown as a
bar (bottom right). The
cross and the filled ellipsoid show the position of Re 4 IRS and HH 188,
respectively. The center of the map corresponds to the position 
$\alpha$(J2000)$=08^h 20^m 58\fs642$ and 
$\delta$(J2000)$=-49^{^\circ} 34' 47\farcs57$.
}
     \label{fhco32} 
     \end{figure}

From the spectra (see Figure~\ref{fhco32} and Table~\ref{tb:obs}) one can see
that the lines are narrow, with a  width of $\Delta v \simeq 0.7$ to 0.8~\kms\
for the \htco\ line, which is optically thin, and  that there are no significant
velocity gradients along the region ($\la 0.4$~\kms). The \hco\ \J{3}{2}\
spectra do not show self-absorption. Redshifted self-absorption in optically
thick lines can be a signature of infall motions (e.g., Zhou et al.
\cite{Zhou93};  Rawlings \& Yates \cite{Rawlings01}).

%
%

The column densities were derived in the following manner: for the SO and \met\
we assumed optically thin emission and an excitation temperature of 10~K.  For
the \hco, we first convolved the \hco\ \J{3}{2}\ emission with a Gaussian to
obtain a final beam size of $57''$, the angular resolution of the \htco\
spectra.  Then, the \hco\ \J{3}{2}\ and \htco\ \J{1}{0}\ optical depth were 
derived from the line ratio of the \htco\ to the convolved \hco\ spectra at the
positions given in Table~\ref{tb:obs}  and assuming a $^{12}$C to $^{13}$C
ratio of 63 (Langer \& Penzias  \cite{Langer93}).   An excitation temperature 
of 7~K was adopted for the \hco\ lines, which is the typical value found for
this molecule in the illuminated condensations ahead of HH objects   (Girart et
al. \cite{Girart02}; Viti et al. \cite{Viti06}).   We find that the \htco\ 
line is optically thin,  $\tau \sim 0.5$, whereas the \hco\ line optically
thick,  $\tau \sim 20$. The column density of the \hco\ is derived from both
the original  and the convolved \hco\ \J{3}{2}\ spectra,  assuming LTE for the
aforementioned excitation temperature and correcting for the optical depth (see
Table~\ref{tb:abunda}).  

In order to derive the fractional abundances we first estimate the total
molecular hydrogen column density from the dust emission. We used the dust
measured with the SEST at 1.3~mm (therefore at a similar angular resolution of
our SO and \hco\ observations) by Reipurth et al.  (\cite{Reipurth93}),
$S_{\rm 1.3mm}=0.27$~Jy. By adopting a dust opacity of $\kappa_{\rm
1.3mm}=1.0$~cm$^2$~g$^{-1}$ (Ossenkopf \& Henning \cite{Ossenkopf94}), a
gas--to--mass ratio of 100 and a dust temperature of 35~K (Reipurth et al.
\cite{Reipurth93}), the beam  averaged column density of the molecular
hydrogen is $2.6\times10^{22}$~\cmd.  Thus, the fractional abundances of SO
and \hco are $X[$SO$]=8\times10^{-10}$ and $X$[\hco]$=4\times10^{-9}$. For the
\met, we first assume that the \met\ and \hco\ trace the same gas. In this
case, the column density ratio of these two species will be the same at 57$''$
(the angular resolution of the \met\ spectra) and at $20''$. From here, the
derived fractional abundance is $X$[\met]$=3\times10^{-9}$.

%
     \begin{table}
     \caption[]{Column density and fractional abundance$^{a}$ with respect to
molecular hydrogen.}
     \label{tb:abunda}
     \[
     \begin{tabular}{lcccc}
     \noalign{\smallskip}
     \hline
     \noalign{\smallskip}   
\multicolumn{1}{l}{Molecule} &
\multicolumn{1}{c}{$N[mol]^b$} &
\multicolumn{1}{c}{$N[mol]^c$} &
\multicolumn{1}{c}{$X[mol]$} 
\\
     \noalign{\smallskip}     
     \hline
     \noalign{\smallskip}
\multicolumn{3}{l}{($0'',0''$)} \\
\hco  & 5.6(13) & 1.1(14)  & 4(-9) \\
\met  & 4.3(13)& \nodata   & 3(-9)  \\
SO    & \nodata& 2.1(13)   & 8(-10) \\
\multicolumn{3}{l}{($0'',-20''$)} \\
\hco  & 5.7(13)  & 1.2(14) & \nodata \\ 
\multicolumn{3}{l}{($20'',-20''$)} \\
\hco  & 4.9(13)  & 1.2(14) & \nodata \\
\multicolumn{3}{l}{($20'',0''$)} \\
\hco  & 4.4(13)  & 7.5(13) & \nodata \\
     \noalign{\smallskip}
     \hline
     \end{tabular}
     \]
     \begin{list}{}{}
\item[$^a$] 
$a(b)$ stands for $a \times 10^b$. Column density units in \cmt. Abundances
derived assuming $N($H$_2)=2.6\times10^{22}$ \cmt.
\item[$^b$] Beam averaged column density for a beam of $57''$ 
\item[$^c$] Beam averaged column density for a beam of $20''$ 
     \end{list}
    \end{table}
%

\section{Inspection of the H$\alpha$ SuperCosmos survey}

\begin{figure} 
     \resizebox{8cm}{!}{\includegraphics{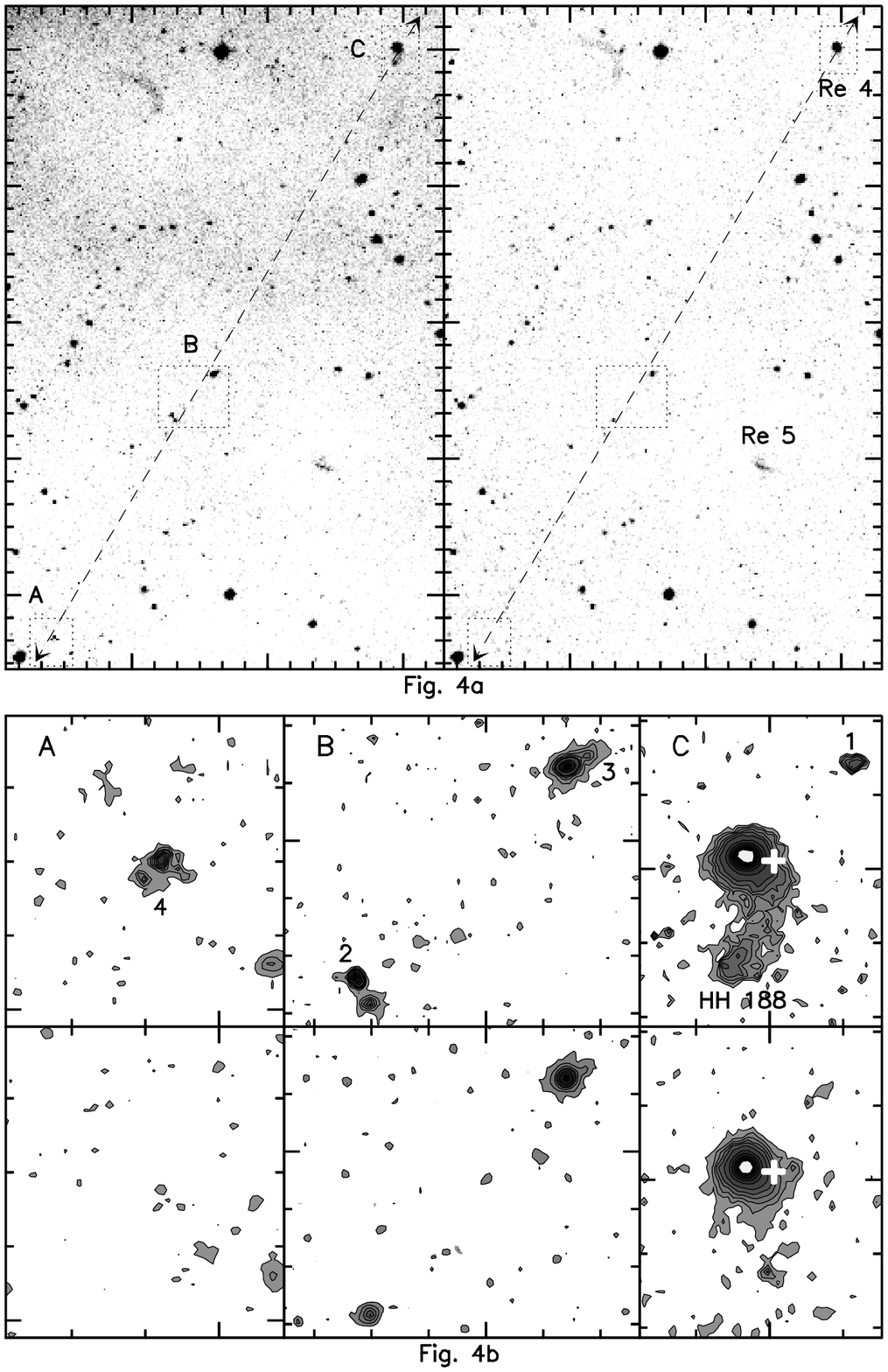}}
     \hfill
     \caption[]{     
     {\em 4(a)}
Narrow--band H$\alpha$ (left panel) and contemporaneous matching 
broad--band Short--Red (right panel) images of the HH~188 region.
The images are obtained from the AAO/UKST SuperCOSMOS 
H$\alpha$ survey (Parker et al. \cite{Parker05}). The dashed
arrowed line shows the direction of the possible HH~188 outflow system.
The small dotted rectangles show the enlarged regions shown in the
bottom panels.
     {\em 4(b)}
Narrow--band H$\alpha$ (top panels) and broad--band Short--Red 
(bottom panels) images of the selected regions in Fig. 4(a),
where pure H$\alpha$ is detected along the direction 
of the HH 188 outflow system. The thick cross in the right panels indicate the
position of the nea-IR peak emission associated with Re~4 IRS.
}
     \label{fhalfa} 
     \end{figure}

We have examined the  H$\alpha$ SuperCosmos survey in a region of
$20'\times20'$ around HH 188 in order to check for possible HH--like objects
that may be associated to this region.   Figure~\ref{fhalfa} shows the narrow
band image centered  at the  H$\alpha$ image and the red broad band continuum
image. We have found several knots of pure H$\alpha$ emission.  Table~\ref{thh}
shows the position, distance and the position angle with respect to  Re~4 IRS
(see \S~\ref{Syso}).  Around Re~4 IRS  there is a compact  (a size of $\sim
3''$) counterpart of HH~188 northwest of the  infrared source.  This knot and
HH 188 are  separated by $31\farcs6$ with a position angle of 148.9$\arcdeg$. 
Southeast of HH~188 we have found three other relatively compact knots  
($\sim3''$) of pure H$\alpha$ emission. Object 2 is close to a  star
($4\farcs5$). However, an inspection of the 2MASS catalogue shows  that this
star has colors, $J-H=0.64$ and $H-K=0.32$, indicating that it is a main
sequence star. This suggests that the H$\alpha$ emission  is not related with
the star.  The five H$\alpha$ objects, including HH 188, appear  very well
aligned (see Fig.~\ref{fhalfa}) with a position angle of $\simeq 149\arcdeg$.
We checked the 2MASS catalogue  and neither of these sources have emission in
the near-IR. The lack of red continuum and near-IR emission suggests that these
objects are probably HH objects, apparently forming part of the same system and
associated with HH~188.

%
%

%
     \begin{table}
     \caption[]{Tentative identification of HH objects}
     \label{thh}
     \[
\begin{tabular}{c@{\hspace{0.1cm}}rr@{\hspace{0.1cm}}rr@{\hspace{0.1cm}}r}
     \noalign{\smallskip}
     \hline
     \noalign{\smallskip}   
\multicolumn{1}{c}{Object} &
\multicolumn{1}{c}{$\alpha$(J2000)} &
\multicolumn{1}{c}{$\delta$(J2000)} &
\multicolumn{1}{c}{Distance$^a$}&
\multicolumn{1}{c}{PA$^a$} &
\multicolumn{1}{c}{} 
\\
\multicolumn{1}{c}{Num.} &
\multicolumn{1}{c}{(08$^h$)} &
\multicolumn{1}{c}{($-49\arcdeg$)} &
\multicolumn{1}{c}{($''$)}&
\multicolumn{1}{c}{($\arcdeg$)} &
\multicolumn{1}{c}{Notes} 
\\
     \noalign{\smallskip}
     \hline
     \noalign{\smallskip}     
1 & 20$^m$57$\fs$10 & 34$'$31$\farcs$2 &  15.8 & 148.7 & \\
2 & 20$^m$58$\fs$78 & 34$'$58$\farcs$1 &  15.8 & 148.7 & HH 188 \\
3 & 21$^m$14$\fs$68 & 39$'$32$\farcs$5 & 333.3& 149.7 & \\
4 & 21$^m$18$\fs$80 & 40$'$11$\farcs$6 & 385.0& 148.2 & \\
5 & 21$^m$29$\fs$47 & 43$'$28$\farcs$5 & 607.0& 149.7 & \\
     \noalign{\smallskip}
     \hline
     \end{tabular}
     \]
     \begin{list}{}{}
     \item[$^{a}$] Distance and position angle with respect to the 
driving source, Re~4 IRS, which is assumed to lie between HH-like
object 1 and HH 188 (see \S~\ref{Syso}).
     \end{list}
    \end{table}
%

\section{Discussion}

\subsection{Re~4 IRS  and the HH~188 system\label{Syso}}

\begin{figure} 
     \resizebox{7.5cm}{!}{\includegraphics{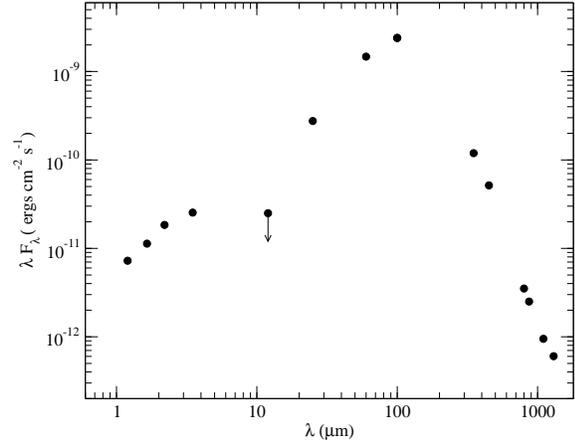}}
     \hfill
     \caption[]{    
Spectral energy distribution of Re 4 IRS, \iras. 
Values at the $JHKL$ near-IR bands 
(1.2, 1.65, 2.2 and 3.4 $\mu$m,respectively) are from Graham \cite{Graham86}.
The values from IRAS bands (12, 24, 60 and 100 $\mu$m) are from 
Cohen \& Schwartz \cite{Cohen87}. Submm and mm values at 350, 450, 
800 and 1100~$\mu$m are from Dent et al. \cite{Dent98} and at 870 and 
1300~$\mu$m from Reipurth et al.  \cite{Reipurth93}.
}
     \label{fsed} 
     \end{figure}

Figure~\ref{fsed} shows the spectral energy distribution (SED) of the driving
source of HH~188, Re~4 IRS.  
 
\begin{figure} 
     \resizebox{7.5cm}{!}{\includegraphics{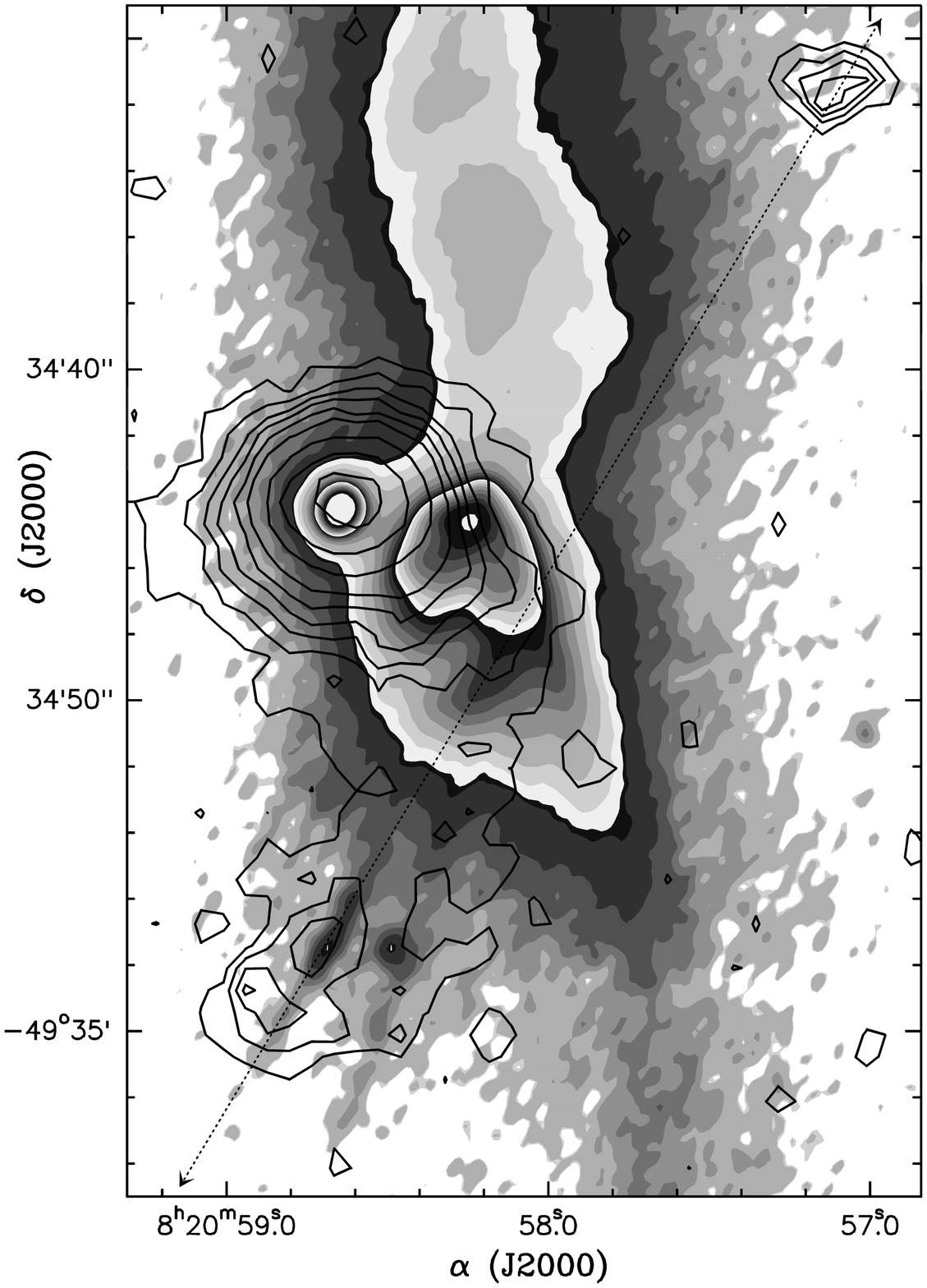}}
     \hfill
     \caption[]{    
Overlap of the contour H$\alpha$ image from the SuperCOSMOS 
H$\alpha$ survey (Parker et al. \cite{Parker05}) and the grey scale
VLT Ks image from Zinnecker et al.  (\cite{Zinnecker99}). In order to 
better show the two faint jets, the Ks image shown here was obtained
by convolving the original VLT image with a Gaussian of 
$0\farcs5\times0\farcs2$ and $PA=149\arcdeg$.
The dotted line shows the direction of the proposed HH system 
associated with HH 188.
}
     \label{fHaKs} 
     \end{figure}

Whitney et al. (\cite{Whitney03}) carried out detailed radiative transfer
models of protostellar envelopes, showing the expected SED and the near-IR
morphology for different inclination angles and evolutionary stages. By using
the Re~4 IRS's SED and the color composite $JHK$  image obtained by Zinnecker
et al. (\cite{Zinnecker99}) we found that this  YSO is possibly a late Class 0
or  Class I object, with an inclination angle of roughly  $\sim 30\arcdeg$.  
Graham \& Heyer (\cite{Graham89}) pointed out that the near-IR  emission
associated with Re~4 IRS has a cometary structure  (see  Fig.~\ref{fHaKs}),
which indicates that is tracing the scattered stellar  light arising
from the cavity created by the outflow. Comparing the images  at $R$, $J$, $H$,
$Ks$, the cometary  structure appears to be displaced to the northwest at
increasing wavelength  (Graham  \& Heyer \cite{Graham89}).  Taking into account
that at increasing  wavelengths the emission arises closer to the YSO, then 
the most reliable position of the source is from the Ks band: from the
Zinnecker et al.  (\cite{Zinnecker99}) image, the position of Re~4 IRS is 
$\alpha$(J2000)$=08^h 20^m 58\fs24$ and  
$\delta$(J2000)$=-49^{^\circ} 34' 44\farcs6$ (with a position uncertainty of 
$\la 0\farcs3$: Correia, private communication).

Figure~\ref{fHaKs} shows that the near-IR emission associated with Re~4 IRS 
is not perfectly aligned with HH~188 and the H$\alpha$ knot 1: these two form 
a position angle of  149$\arcdeg$, whereas the Ks emission  peak of Re~4 IRS
has position angle of  158$\arcdeg$ and 140$\arcdeg$ with HH~188 and the
H$\alpha$ knot 1, respectively. This suggests that the  powering source of the
HH 188 system should be located roughly few arcsecs west of near-IR  peak
emission.  A tentative position for the powering source can be  obtained 
assuming that it is located equidistantly between HH~188 and the H$\alpha$ 1, 
that is: 
$\alpha$(J2000)$=08^h 20^m 57\fs94$ and 
$\delta$(J2000)$=-49^{^\circ} 34' 44\farcs7$. This is $\sim3''$ west of the 
Re~4 IRS position derived previously from the Ks image. Table~\ref{thh} shows
the distance and position angle of the different  H$\alpha$ knots with respect
to this tentative position: all the  H$\alpha$ knots are very well 
aligned with respect to this position with a position angle of 
$149\arcdeg\pm1\arcdeg$. In addition, as can be seen in Figure~\ref{fHaKs}, one
of the two jets detected by Zinnecker et al. (\cite{Zinnecker99}) coincides 
with HH~188 and is also well aligned with the  direction of the HH~188 system.
Thus, as already suggested by Zinnecker et al.   (\cite{Zinnecker99}), it seems
that Re~4 IRS is a binary system.  The protostar powering the HH~188 system is
possibly a very embedded object, too faint in the near-IR to be detected. On
the other hand, the protostar  associated with the near-IR cometary-like
emission could be the powering   source of the other jet detected in the Ks
image (see Fig.~\ref{fHaKs}).

The different H$\alpha$ knots aligned with HH 188 and Re~4 IRS span over
a length of 10$'$ or 1.2 pc in projection. The two brightest spots are the two 
farthest and southernmost objects, 4 and 5 from Table~\ref{thh}. The properties
of the HH 188 can be derived from the spectroscopical
observations obtained by Graham (\cite{Graham86}): Table 5 of that paper gives
the relative intensities (with respect to H$\alpha$) of the different emission
lines detected ($[$SII$]$, H$\alpha$, $[$OI$]$ and $[$NaII$]$).  From the 
$\lambda6716+\lambda6730[$SII$]$/H$\alpha$ and $[$SII$]$
$\lambda6716$/$[$SII$]$ $\lambda6730$  lines ratios we estimate that HH~188 is
a high excitation object with a high electron density
(n(e$^-$)$\simeq$2000~cm$^{-3}$). For the other knots further observations are 
required to characterize their properties.

%
%
%
%

\subsection{The origin of the \hco\ emission}

The \hco\ 3--2 peak is  close to HH~188 and offset by $\simeq 18''$ to the 
southeast with respect to the peak position of Ks emission from  Re~4 IRS.
If, instead we use the tentative position of the HH~188 system's driving source,  
the offset is $\simeq 20''$. This offset between the HCO$^+$ 
emission and the protostar is comparable to the beam diameter of the  HCO$^+$ 
map, so it is significative. 

The fractional abundances derived in Section 3 {\it assume} that the emission
comes from the central source, as we derived the H$_2$ column density from the
dust measurement at 1.3 mm that peaks at Re~4 IRS. However, the properties of
the molecular emission that we see i.e (i) a displacement to the southeast
close to HH~188, (ii) no significant velocity gradients, and (iii) narrow line
widths, suggest  that the molecular gas comes,  at least partly,  from an
unperturbed dynamically or quiescent region.  Nevertheless, because the angular
resolution of our observations is not high enough to spatially resolve the IR
and HH objects, we can not easily determine where the molecular emission comes
from. Hence below we qualitatively investigate different possibilities.

At first we consider the most straightforward possibility that the emission
comes from the dense circumstellar envelope around the protostar Re~4 IRS.  
If we assume that the SO and \hco\ come from the cold, outer envelope  around
the protostar, then their abundances are a factor of $\sim 2$ respectively
lower and higher with respect to the average value in the sample of Class 0
and I low mass protostars studied by J{\o}rgensen \et\  al.
(\cite{Jorgensen04}), but still within the range of derived values.  The \met\
is expected to be significantly enhanced in the warm, inner  envelopes ($T\ga
90$~K) around low mass protostars (Maret \et\  \cite{Maret05}; J{\o}rgensen
\et\ \cite{Jorgensen05}).  Our observations of \met\ around Re~4 IRS yield an
abundance higher  than the range of values for the outer envelopes given by
Maret \et\  (\cite{Maret05}) but much lower than the values found for the
inner  envelopes.  The low \met\ line width ($\simeq 0.6$~\kms)  and the
non-detection of the \JEo\ transition  in Re~4 IRS discards the inner envelope
origin: in the Maret \et\ (\cite{Maret05})  sample, the line widths are always
larger than 1~\kms, with several cases where  the line widths are few \kms,
and the \JEo\ transition should have been detected for densities higher than
few times $10^5$~\cmt.    Therefore, our observations suggests that a
significant fraction of the \met\ emission does not arise from the envelope
around the IRAS protostar. In fact in   other star forming regions, quiescent
methanol  emission has been detected significantly offset from the protostars
and  not too far from shocks (Palau \et\ \cite{Palau07}; Girart \et\
\cite{Girart07}).

These concerns led us to take into consideration alternative scenarios for the
origin of at least a significant fraction of the observed molecular  emission. 
It could be that most of the molecular emission arises from the interface along
the HH~188 jet with the ambient cloud: in low-mass star forming regions,
observations of enhanced HCO$^+$ toward several Class 0 objects can indeed be
interpreted as coming from the walls of the outflow cavity (Hogerheijde et al.
\cite{Hogerheijde98}; Rawlings et al. \cite{Rawlings00}; Redman et al.
\cite{Redman04}). It is not unreasonable to assume that highly turbulent
interfaces may be present in all environments where HH jets interact with the
surrounding molecular cloud. Such interfaces have been shown to be
characterized by a chemistry that is quite distinct in nature from that of
typical dense cores and would not be achieved by any modification of
conventional cold-cloud chemistry (Lim et al. \cite{Lim01}; Rawlings \&
Hartquist  \cite{Rawlings97}). 
However, if the emission were coming from an interface one
would expect larger ($\ge$ 1 km s$^{-1}$) linewidths.

%
     \begin{table}
     \caption[]{Abundance ratios$^a$}
     \label{tb:abunda2}
     \[
     \begin{tabular}{lccc}
     \noalign{\smallskip}
     \hline
     \noalign{\smallskip}   
\multicolumn{1}{l}{Region} &
\multicolumn{1}{c}{$X[$SO$]/X[$HCO$^+]$} &
\multicolumn{1}{c}{$X[$\met$]/X[$HCO$^+]$} 
\\
     \noalign{\smallskip}     
     \hline
     \noalign{\smallskip}
HH~188 & 0.2  & 0.8   \\
HH~2~I1  & 0.2 & 1.2   \\
HH~34E  & 0.9 & 1.5  \\
HH~1 & 0.8 & 2.1 \\
     \noalign{\smallskip}
     \hline
     \end{tabular}
     \]
     \begin{list}{}{}
\item[$^a$]  The uncertainty of these ratios is approximately 30\%
     \end{list}
    \end{table}
%

Another possibility is whether the emission arises from the cold, illuminated,
gas ahead of the HH~188 object: quiescent, cold, molecular condensations have
been found ahead of several HH objects in enhanced emission of HCO$^+$ (e.g. 
Rudolph \& Welch \cite{Rudolph88}; Torrelles et al. \cite{Torrelles92}; Girart
et al. \cite{Girart05}) and can be  explained as the result of photochemistry
generated by UV radiation from  the HH shock which leads to evaporation of icy
mantles on dust grains in  small density enhancements (e.g. Viti \& Williams
\cite{Viti99}). 
In favour of this scenario it is important  to point out that the 
$\lambda6716+\lambda6730[$SII$]$/H$\alpha$  lines  ratio estimated in HH~188 is
similar to the one derived for HH~34  (Morse et al. \cite{Morse92}). HH~34, as
well as other high-excitation objects,  is known to have apparently
illuminated dense molecular gas. Recently Viti et al. (\cite{Viti06}) presented
a molecular survey ahead of several HH objects and found that the typical
density, temperature, and line widths of such clumps are 10$^5$-10$^6$
cm$^{-2}$,  10-20~K, and 0.4-1.5~\kms\ respectively; the derived ranges of
column densities  for HCO$^+$, \met\ and SO were respectively:
5$\times$10$^{12}$--2$\times$10$^{14}$,
3$\times$10$^{13}$--4$\times$10$^{14}$, and
8$\times$10$^{12}$--9$\times$10$^{13}$\cmd. The observed column densities
around \iras\ are indeed consistent with these ranges. This together with the
fact that the 
HCO$^+$ peak is displaced with respect to the IRS source and the
line widths of the emission lines are of the same order as those from the
molecular emission detected ahead of all the HH objects in the survey make this
scenario a plausible one.  It is then interesting to compare the ratio of
column densities derived here with those from the objects in the survey by Viti
et al. (\cite{Viti06}) and with HH~2, whose complex interaction with its
surroundings has been extensively studied (Girart et al. \cite{Girart02};  Viti
et al. \cite{Viti03}; Lefloch et al. \cite{Lefloch05}; Girart et al.
\cite{Girart05}).  We list these ratios in Table~\ref{tb:abunda2}.  The reason
for using ratios rather than individual column densities is to make the
comparisons independent of the telescope.  Both ratios indicate that if the
molecular emission is indeed coming from the quiescent gas ahead of the HH
object then it is closer in characteristics to HH~2 than HH~1 and HH~34. The
region surrounding HH~2 is very complex: for this comparison we chose to use
the ratios derived from the densest (and closer to the HH~2) part of the region
that Girart et al. (2005) interpret as being quiescent and directly illuminated
by the UV coming from HH~2 (measured to be $\sim$40 Habing, see Molinari \&
Noriega-Crespo \cite{Molinari02}). Hence, if this scenario is correct, the gas
we observe around HH~188 has a density of $\sim$ 10$^5$ cm$^{-3}$, and a
temperature of $\sim 15$~K.

Finally, we consider the possibility that the background radiation from the Gum
Nebula is the source of the UV radiation that is altering chemically the
molecular  gas in a similar way as the HH object.  The Gum Nebula shows diffuse
and extended   H$\alpha$ and radio emission, which suggests that it has also
associated significant  UV radiation generated by OB stars and/or by and old
supernova remnant  (e.g. Woermann, Gaylard \& Otrupcek \cite{Woermann0}). In
order to consider this possibility we checked the SHASSA  H$\alpha$ survey 
(Gaustad et al.  \cite{Gaustad01}). The Sandqvist 109 molecular cloud (which
includes the Re~4 IRS  region) is at a distance of $\sim30\arcmin$ from the
brightest H$\alpha$ emission  regions of the Gum nebula. The background
H$\alpha$ emission around Re~4 IRS regions is about 2 to 3 times weaker than
these bright regions.  We have also checked the SHASSA survey around two other
high excitation HH objects with irradiated dense molecular clumps, HH~2  and
HH~34, both in the Orion star  forming region, where there is also diffuse
H$\alpha$ emission. The level of  diffuse H$\alpha$ emission around these two
objects is similar to the level in  the Re~4 IRS region.  HH~34 (Viti et al.
\cite{Viti06}) and in particular HH~2 (Girart et al. \cite{Girart05}) exhibit 
molecular emission that has been succesfully modelled using the UV 
radiation--driven chemistry scenario. Although the possibility that the 
diffuse UV radiation contributes to the illumination of the molecular 
condensation can not be ruled out, the close spatial connection of  the
molecular emission with the HH objects indicates that the main  source of UV
radiation comes from the HH objects.

\section{Summary}

In this paper we presented maps of the region around Re~4 IRS
in HCO$^+$ (3-2) and H$^{13}$CO$^+$ (1-0), as well as observations in
SO (6$_5$--5$_4$) and CH$_3$OH (2$_0$-1$_0$ A$^+$) of the central
position.  We find that the HCO$^+$ emission is quite compact around
the protostar but its peak is displaced to the southeast close to the
Herbig-Haro object HH~188. All the lines are narrow ($\sim$ 0.7 
km~s$^{-1}$) and HCO$^+$ does not show any sign of infall. Column
densities were derived assuming LTE conditions.

\par We have also checked the SuperCosmos H$\alpha$ survey and have found
several knots of pure H$\alpha$ emission very well aligned with HH 188 and
forming an apparent parsec scale HH system. The driving source of this system
is possibly an embedded protostar invisible in the near-IR. We tentatively
estimate that it is located about $\sim3''$  west of the Ks emission peak.  The
protostar illuminating the near-IR cometary--like nebula is possibly the 
exciting source of a faint jet. Thus, Re~4 IRS, a late Class 0 or Class I
object,  seems to be a binary system, as suggested by Zinnecker et al.
\cite{Zinnecker99}.  

\par We attempt a very simple chemical analysis of the molecular  emission in
order to shed light on its origin. We consider four scenarios  as the possible
origin of the emission:  the circumstellar envelope around  the protostar; an
interface along the HH~188 jet with the ambient cloud; the quiescent, cold,
illuminated gas ahead of the HH~188 jet;  illumination by the diffuse Gum
Nebula emission.  The derived column densities, as well as the narrowness of
the linewidths  and the morphology of the HCO$^+$ emission are in agreement
with the  third scenario making it a plausible one, although it is still
possible that there is  contribution from the outer cold envelope around the
protostar.

\par These observations have shown that the Re 4 IRS--HH~188 would  make an
interesting southern target for future ALMA studies of star forming regions.  

\begin{acknowledgements}
We thank H. Zinnecker and S. Correia for providing the VLT Ks image. JMG
acknowledges support by MCyT grant AYA2005-08523-C03-02. SV acknowledges
individual financial support from a PPARC Advanced Fellowship.  JMG and SV  
acknowledge support by a joined Royal Society and CSIC travel grant. This
publication makes use of data products from the  Two Micron All Sky Survey,
which is a joint project of the University of  Massachusetts and the Infrared
Processing and Analysis Center/California  Institute of Technology, funded by
the National Aeronautics and Space  Administration and the National Science
Foundation.

\end{acknowledgements}


%
     \begin{table*}
     \caption[]{Molecular line results for HH 188}
     \label{tb:obs}
     \[
     \begin{tabular}{lcccc}
     \noalign{\smallskip}
     \hline
     \noalign{\smallskip}   
\multicolumn{1}{l}{Molecular} &
\multicolumn{1}{c}{$T_{\rm mb}$} &
\multicolumn{1}{c}{$\int T_{\rm mb} dv$} & 
\multicolumn{1}{c}{$v_{\rm LSR}$} &
\multicolumn{1}{c}{$\Delta v_{\rm LSR}$} 
\\
\multicolumn{1}{l}{Transition} &
\multicolumn{1}{c}{$\!\!\!\!\!$(K)} &
\multicolumn{1}{c}{$\!\!\!\!\!$(K\kms)} & 
\multicolumn{1}{c}{$\!\!\!\!\!$(km~s$^{-1}$)} &
\multicolumn{1}{c}{$\!\!\!\!\!$(km~s$^{-1}$)} 
\\
     \noalign{\smallskip}     
     \hline
     \noalign{\smallskip}
\multicolumn{4}{l}{($0'',0''$) offset position} \\
\htco \J{1}{0}\    & 0.80(2) & 0.59(2) & 3.79(1) & 0.70(3) \\
\hco \J{3}{2}\     & 2.32(26)& 2.20(16)& 3.60(3) & 0.89(7) \\
\met\ \JA          & 0.44(5) & 0.27(2) & 3.74(1) & 0.57(2) \\
\met\ \JE          & 0.44(5) & 0.27(2) &fixed$^a$&fixed$^a$ \\
\met\ \JEo         & $\la0.12$ & \nodata &fixed$^a$&fixed$^a$ \\
SO \JK{6}{5}{5}{4}& 0.81(5) & 0.69(3) & 3.64(1) & 0.81(4) \\
\hline
\multicolumn{4}{l}{($0'',-20''$) offset position} \\
\htco\ \J{1}{0}   & 0.83(2)  & 0.65(2) & 3.74(1) & 0.73(3) \\
\hco\ \J{3}{2}    & 2.43(22) & 2.12(13)& 3.52(3) & 0.82(6) \\ 
\hline
\multicolumn{4}{l}{($20'',-20''$) offset position} \\
\htco\ \J{1}{0}   & 0.66(2)  & 0.51(2) & 3.74(2) & 0.72(4) \\
\hco\ \J{3}{2}    & 1.82(28) & 2.00(18)& 3.63(5) & 1.03(10) \\
\hline
\multicolumn{4}{l}{($20'',0''$) offset position} \\
\htco\ \J{1}{0}   & 0.60(4)  & 0.51(3) & 3.85(2) & 0.80(4) \\
\hco\ \J{3}{2}    & 1.65(26) & 1.68(19)& 3.64(5) & 0.95(14) \\
     \noalign{\smallskip}
     \hline
     \end{tabular}
     \]
     \begin{list}{}{}
\item[$^a$] 
The three methanol lines were fitted simultaneously 
     \end{list}
    \end{table*}
%
%

\end{document}